# Investigating DNA words and their distributions across the tree of life

Charalampos Koilakos[1], Kimonas Provatas[1], Michail Patsakis[1], Aris Karatzikos[1], Ilias Georgakopoulos-Soares[1,*]

[1] Division of Pharmacology and Toxicology, College of Pharmacy, The University of Texas at Austin, Dell Paediatric Research Institute, Austin, TX, USA.
[*] Corresponding authors: izg5139@psu.edu

## Abstract

The frequency distributions of DNA k-mers are shaped by fundamental biological processes and offer a window into genome structure and evolution. Inspired by analogies to natural language, prior studies have attempted to model genomic k-mer usage using Zipf's law, a rank-frequency law originally formulated for words in human language. However, the extent to which this law accurately captures the distribution of k-mers across diverse species remains unclear. Here, we systematically analyze k-mer frequency spectra across more than 225,000 genome assemblies spanning all three domains of life and viruses. We demonstrate that Zipf's law consistently underperforms in modeling k-mer distributions. In contrast, we propose the truncated power law and Zipf-Mandelbrot distributions, which provide substantially improved fits across taxonomic groups. We show that genome size and GC content influence model performance, with larger and GC-content imbalanced genomes yielding better adherence. Additionally, we perform an extensive analysis on vocabulary expansion and exhaustion across the same organisms using Heaps' law. We apply our modeling framework to evaluate simulated genomes generated by k-let preserving shuffling and deep generative language models. Our results reveal substantial differences between organismal genomes and their synthetic or shuffled counterparts, offering a novel approach to benchmark the biological plausibility of artificial genomes. Collectively, this work establishes new standards for modeling genomic k-mer distributions and provides insights relevant to synthetic biology, and evolutionary sequence analysis.

## Introduction

Genome sizes exhibit remarkable variation across the tree of life, spanning from the minimalistic genomes of certain viruses to the expansive genomes of certain eukaryotes. In particular, some viruses, such as Porcine circovirus type 1, have genomes as small as 1,766 nucleotides [1]. In contrast, certain plant species exhibit genome sizes exceeding 100 gigabase pairs (Gbp), with the *Paris japonica* (a flowering plant) having a genome of approximately 150 Gbp [2]. K-mers are short, contiguous nucleotide sequences of length k, and have a number of biological applications [3]. The k-mer profile of an assembled genome can be modeled based on the frequency of its constituent k-mers [4,5]. Analyses of k-mer profiles across different genomes show that most species have a unimodal k-mer spectrum; however certain species including mammals, have multimodal spectra, which can be explained by differences in the mono- and di-nucleotide background frequencies of these species [5].

The structure and evolution of genomic sequences exhibit certain statistical patterns reminiscent of linguistic systems. These principles, observed in natural languages, could also capture the distribution and organization of certain genomic elements [6,7]. Power-law distributions have been widely applied in biological modeling, including in the analysis of networks such as protein-protein interactions and metabolic pathways [7,8]. Zipf's law is an empirical principle that describes the frequency distribution of elements in a dataset. It states that the frequency of an item is inversely proportional to its frequency rank. Zipf's law has found multiple applications in genomics and molecular biology. For example, one study showed that gene expression exhibits a power-law distribution that follows Zipf's law [9]. Similarly, Zipf's law was implemented to study k-mer frequencies in organismal genomes [10]. However, later works indicated that this relationship was not evident and there was a poor power-law relationship [11–13]. One study demonstrated that selfish DNA elements primarily drive the power-law tail of the k-mer frequency distribution [14]. A recent study analyzed k-mer frequencies in 178 genomes and found that both their Zipf-like distributions and inverse symmetries can be explained probabilistically by the Conservation of Hartley-Shannon Information [15].

Another linguistic law is the Menzerath-Altmann law, which states that the longer a construct, the shorter its constituents. In genomics, Menzerath's law is reflected in the observation that genes with more exons tend to have shorter individual exons [16]. Heaps' law describes the sublinear growth of the number of unique elements in a dataset as its size increases, and it has been applied extensively in linguistics [17]. One study showed that Heaps' law can be used to model the openness of a pangenome [18,19]. Another study examined the applicability of Heaps' law to genomic sequences, using protein domain-coding regions as DNA words [20]. These findings support the development of mathematical models for k-mer frequency analyses.

Here, we examined multiple distributions to model k-mer frequency profiles of over 225,000 organismal genome assemblies. We applied Heaps' law to characterize how k-mer diversity grows with sampling, revealing taxon- and k-specific regimes that reflect size and compositional differences between organismal genomes. We also report that Zipf's law consistently underperforms and does not capture the frequency-rank relationship of k-mer frequencies. We propose the usage of the Zipf-Mandelbrot distribution or the truncated power law to model k-mer frequency-rank relationships of organismal genomes. We also observe that the fit of frequency-rank models varies by taxonomy, due to genome size variations, and there is an optimal k-mer length fit across most groups. Finally, we show that we can test the realism of shuffled and synthetic genomes by applying the Zipf-Mandelbrot and truncated power laws and fitting their parameters to the genome, allowing direct comparison with the empirical distribution of k-mer frequencies in real genomes.

## Results

### Global patterns of k-mer frequency distributions across organismal genomes

Leveraging an extensive dataset comprising more than 225,000 genome assemblies, we systematically characterized global patterns of k-mer usage and assessed how they vary across taxonomic groups and k-mer lengths. Across the organismal genome assemblies that we analyzed, we find that k-mer distributions follow a power-law decay, which is consistent across the three superkingdoms and viruses (**Figure 1a-d**). Eukaryotic genomes exhibit the most gradual decay in k-mer frequencies, due to the larger genome sizes, while viral genomes display sharper drop-offs and shorter rank ranges. Additionally, in higher k-mer lengths we observe more abrupt cutoffs, indicating that not all possible k-mers of that length are present in every species.

To estimate the degree of unequal k-mer distribution in the organismal genome assemblies we utilized Lorenz curves. The Lorenz curve for k-mers reveals a highly unequal distribution of k-mer frequencies, with a small fraction of k-mers accounting for the majority of total occurrences (**Figure 1e; Supplementary Figure 1**). This inequality becomes more pronounced with increasing k-mer length, reflecting the sparsity and dominance of repetitive sequences at higher k-mer lengths. To complement our findings we also estimated the Gini coefficient, which represents the surface area between the Lorenz curve and the identity line ($y=x$). The Gini coefficient results show a consistent increase with k-mer length across all taxonomic groups, indicating that k-mer frequency distributions become progressively more unequal as k-mer length increases (**Figure 1f; Supplementary Figure 2**). This reflects a shift from a more balanced usage in the case of short k-mers, which are widespread due to overlap and sequence redundancy, to a highly skewed distribution at higher k-mer lengths, where most k-mers are rare or unique and only a few dominate, likely due to repetitive genomic regions and functional motifs.

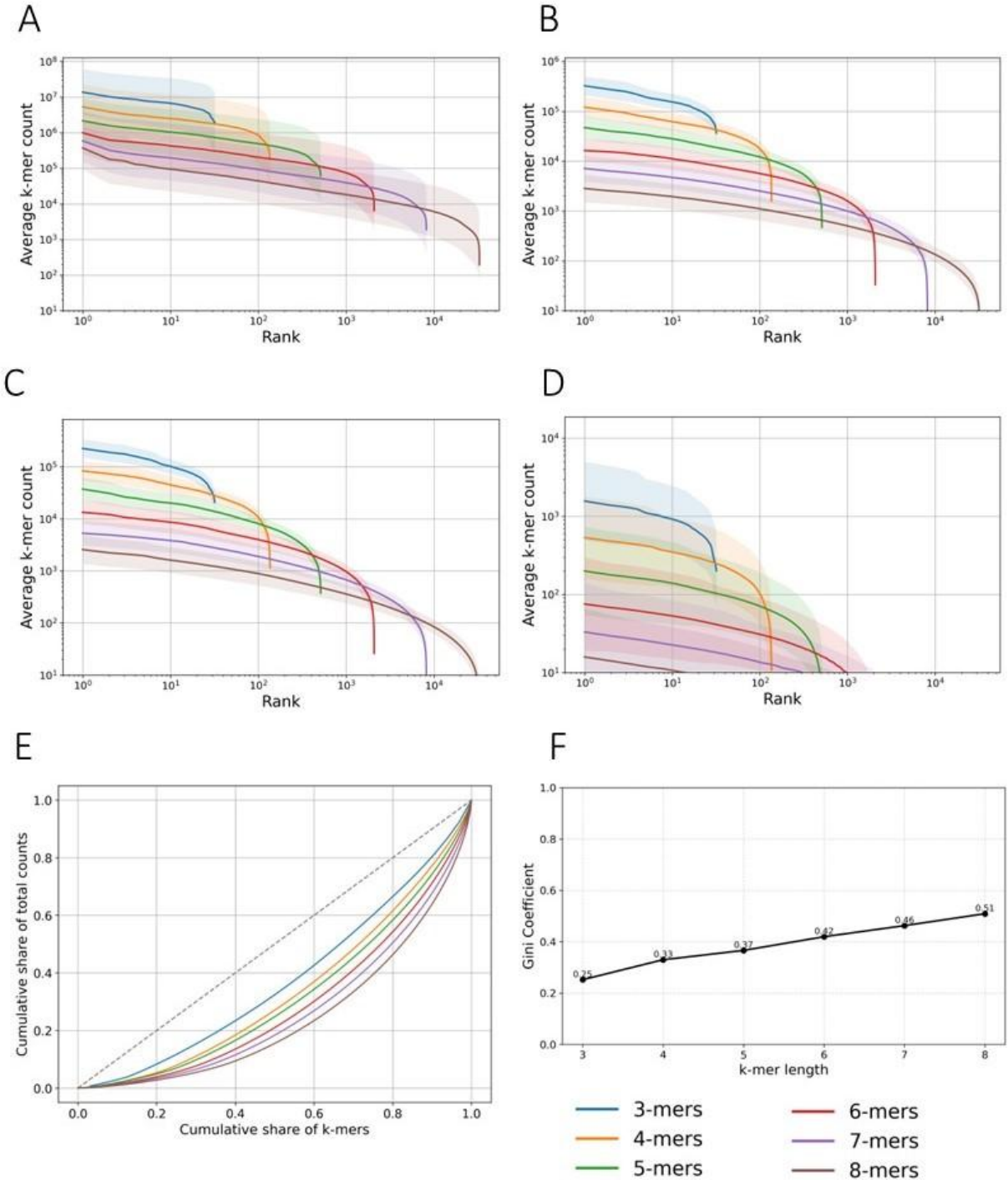

**Figure 1. Global patterns of k-mer frequency distributions across organismal genomes.**
**(A–D)** Average k-mer frequency versus rank plotted on a log-log scale for each major taxonomic group: **(A)** Eukaryota, **(B)** Bacteria, **(C)** Archaea, and **(D)** Viruses. Each curve represents the mean distribution across thousands of genomes at a given k-mer length. As k increases, the distributions become more skewed, with a sharper decline in frequency for lower-rank k-mers and an abrupt cutoff reflecting absent high-order k-mers. **(E)** Lorenz curves illustrating the inequality of k-mer distributions across taxonomic groups. **(F)** Gini coefficients derived from the Lorenz curves quantify k-mer usage inequality, with higher k values showing increased inequality due to vocabulary expansion and sparsity of high-order k-mers in genomes.

**Estimating vocabulary growth from k-mer sampling with Heaps' law**

Heaps' law is an empirical law that describes how the number of unique words in a corpus grows as more words are added. It is formulated as : $V(n) = Kn^\beta$ where n is the total tokens up to a sampling point, V are the distinct tokens (words) you have encountered so far, K is a corpus dependent scale parameter and β is an exponent that signals how fast the vocabulary grows and depends on the language and register. For English texts, β is around 0.4 to 0.6 [21]. When β = 0, the number of distinct words remains constant regardless of corpus size, new words are exceedingly rare, and a small sample is sufficient to capture the entire vocabulary. This typically occurs in systems with highly constrained vocabularies. In genome analyses, this behavior is observed for k-mers with small k values, where the total number of possible k-mers is limited. At the other extreme, when β = 1 or approaches 1, nearly every token encountered is unique, and the stream of words shows minimal or no reuse. For Heaps' law to hold, β must be less than 1, indicating sublinear vocabulary growth. In our analysis, we used canonical k-mers and sampled each genome in 1% increments, tracking the ratio of distinct to total k-mers at each sampling point.

For any given genome and k-mer size, we can associate a specific set of Heaps' law parameters (β and K), characterizing the rate and extent of k-mer diversity expansion across the genome. We first applied Heaps' law in the genomes of *Homo sapiens* and *Escherichia coli*. *Homo sapiens* shows a gradual onset of sublinear vocabulary growth starting at k=11 due to its larger genome size (**Figure 2a-b**). *E. coli* starts showing this behavior at k=8 and we observe that at k=15 we have reached a slope of 1 meaning that β=1 and that almost every k-mer observed has not been encountered before.

We next expanded our analysis across the 225,000 assemblies to examine the range of β and K parameter differences between taxonomic groups. A clustering occurs for all tested k-mer sizes (**Figure 2c-d**), with eukaryotes showing the larger $K$ values, as well as the smallest $\beta$ among taxonomic groups. Viral data on the other hand, show the larger $\beta$ and $K$ values, with archaea and bacteria showing intermediate values. This is expected due to the huge differences in genome size that impact vocabulary exhaustion. Among the four superkingdoms, eukaryotes exhibit a marked increase in β beginning at k = 9, which can be attributed to their larger genome sizes. At k = 15, β values approach 1, indicating near-maximal vocabulary expansion. This indicates that for k up to 9 or 10, nearly all possible k-mers are observed after sampling just 1% of the genome. For k between 11 and 14, Heaps' law holds, but at higher k values, β trends toward 1, suggesting a nearly linear vocabulary expansion (**Figure 3**). In both archaea and bacteria, Heaps' law holds for k values between 9 and 13. Due to their smaller genome sizes, they reach β ≈ 1 more rapidly than eukaryotes. Viral data reach β=1 significantly faster and every seen kmer is new from k=10. Heaps' law holds true for k=6 (and potentially earlier) until k=8 (**Figure 3**). We conclude that genome size and complexity shape the k-mer diversity landscape, with each species and each taxonomic group being characterized by unique Heaps' law parameters for a given k-mer length.

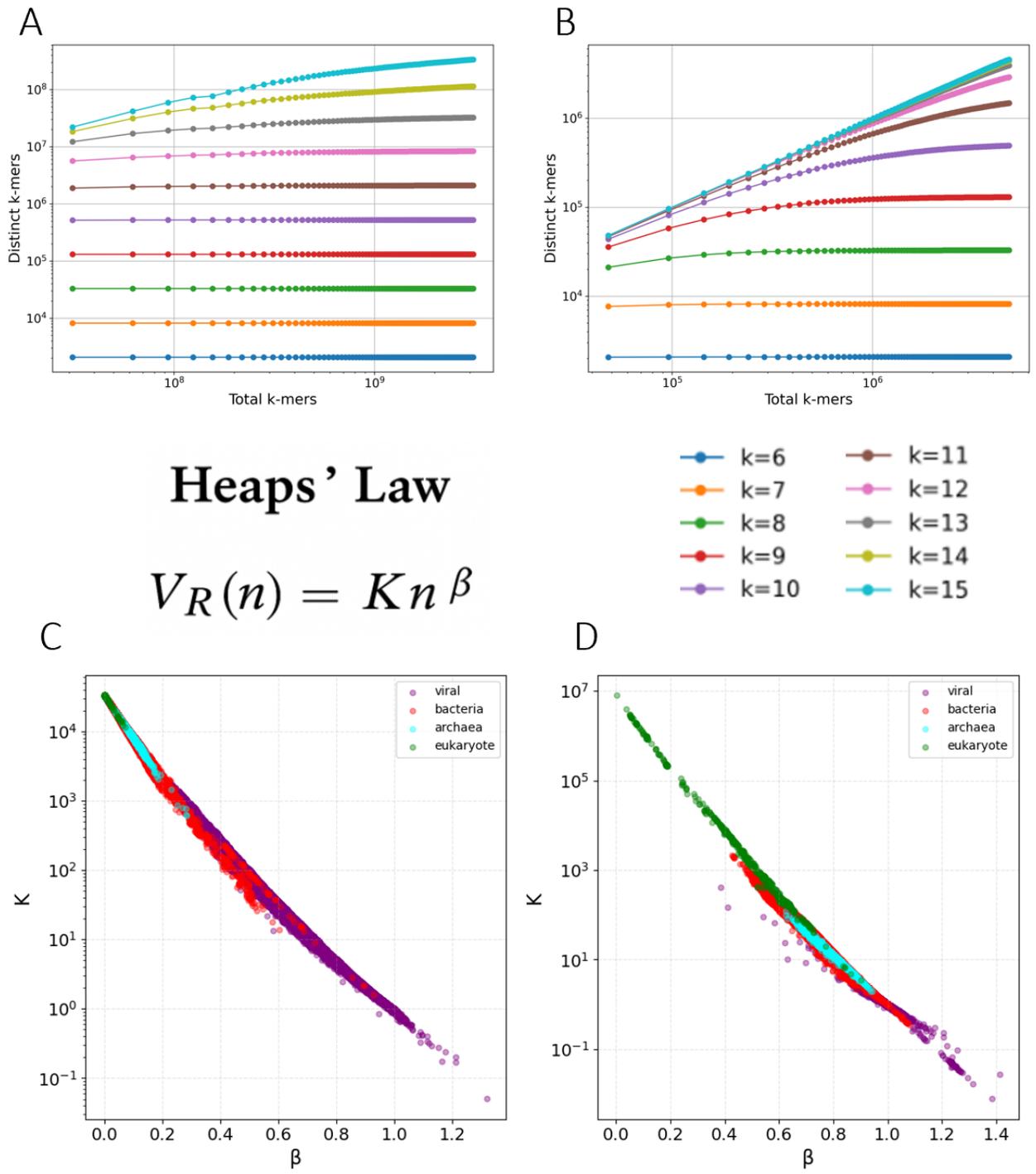

**Figure 2: Estimating genomic vocabulary growth via Heaps' law across species.** Depiction of the relationship between distinct k-mers sampled at every 1% of the genome size and the total k-mers found at the same point for **(A)** *Homo sapiens* and **(B)** *Escherichia coli* in logarithmic scale. Scatter plots illustrating the distribution of K (logarithmic scale) versus β across taxonomic groups for k=8 **(C)** and k=12 **(D)**.

A
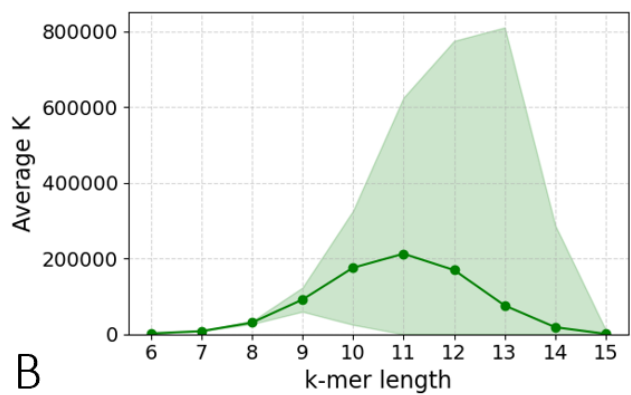
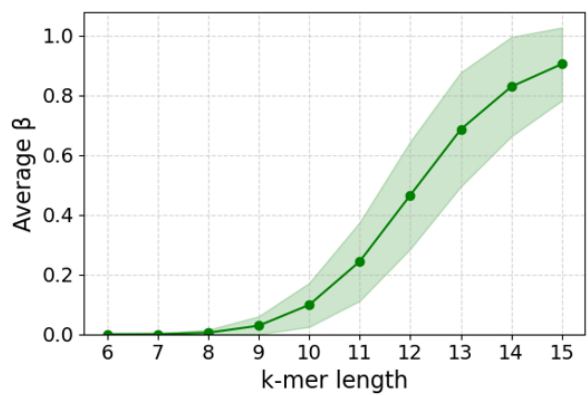

B
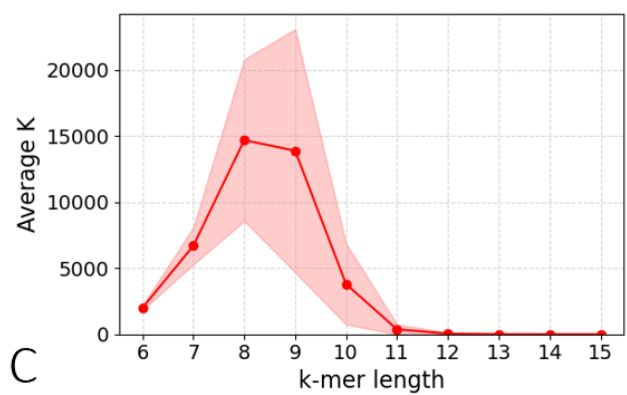
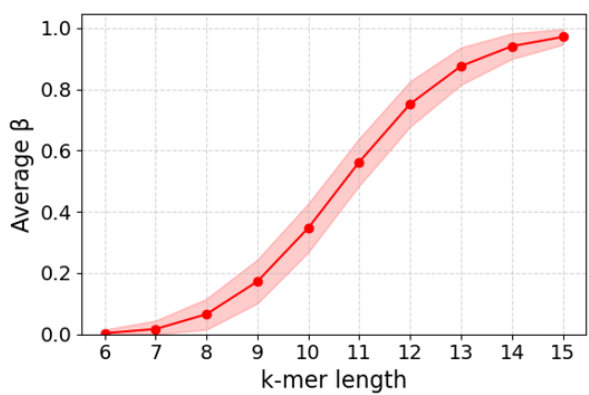

C
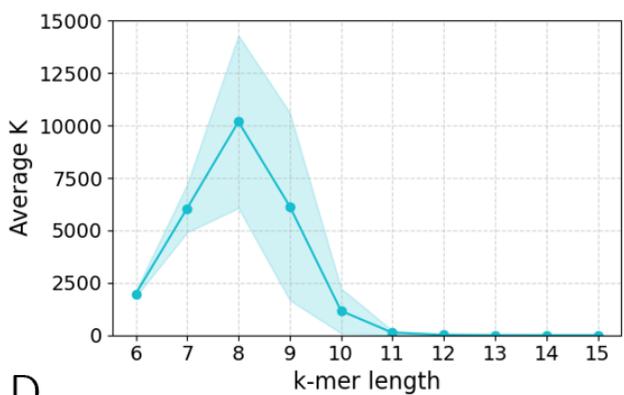
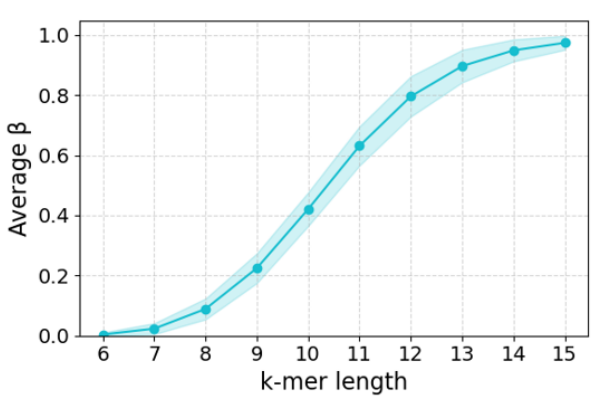

D
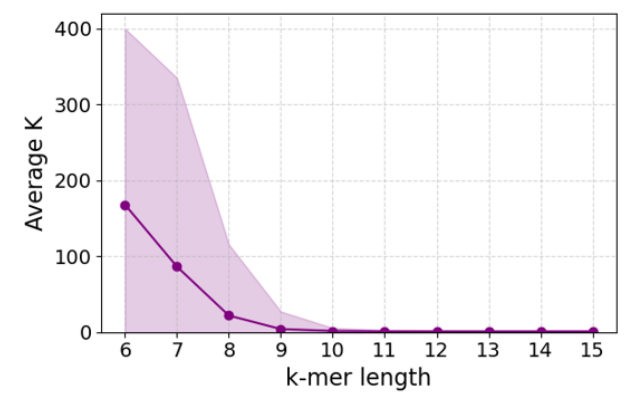
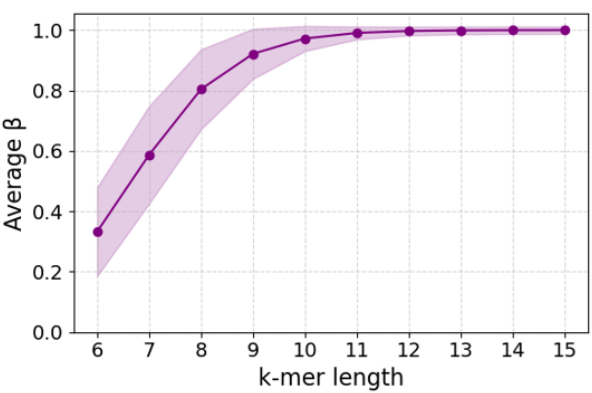

**Figure 3: Distributions of Heaps' law parameters across organismal genomes and taxonomic groups.** Average fitted Heaps' Law parameters K and β for **(A)** Eukaryotes**, (B)** Bacteria**, (C )** Archaea and **(D)** Viruses**. Confidence intervals represent one standard deviation above and below the mean.

**Connection to the Menzerath-Altmann law**

One can draw an instructive parallel between Heaps' law and the Menzerath-Altmann law. The latter states that, in language, the larger a linguistic unit is, the shorter its immediate constituents become. If we treat a genomic DNA sequence as a text, the total number of k-mer tokens $n$ is analogous to text length, while the number of distinct k-mers $V(n)$ plays the role of the vocabulary's constituents. To test whether a Menzerath-Altmann-like effect appears in genomes, we must ask whether genomes that contain more k-mers in total acquire novel k-mers at an ever-slower rate.

By reformulating the Heaps' law equation, $V(n)/n = K * n^{\beta-1}$ and assuming no exponential decay for the Menzerath-Altmann law, $Y = \alpha * X^b$, we observe the similarities. If we consider Y to be the total number of k-mers and X the number of distinct k-mers, we can match $a = K$ and $b = \beta - 1$. Thus by considering that $0 < b < 1$ in sub-linear Heaps' behaviour, we obtain $b < 0$, confirming the inverse relationship that the Menzerath-Altmann law describes. Genomes whose distinct $k$-mer repertoire saturates quickly (small β) exhibit a steeper negative Menzerath slope (more negative b). Conversely, any departure from sub-linearity (β ≥ 1) would algebraically produce $b > 0$, signalling a breakdown of the law at those scales.

In our analysis, we searched across taxonomic groups and k-mer lengths for fits that yield β values approaching zero. Eukaryotes stand out as the average β values are very close to 0 for k-mer lengths up to 11 bps. Beyond this point the values began to deviate further from the Menzerath-Altmann hypothesis (**Figure 3a**). Archaea and Bacteria show similar tendencies with each other, where the Menzerath-Altmann law holds true for k-mer values below 10 bps (**Figure 3b-c**). Viral genomes on the other hand present the largest deviation from our hypothesis where we have to search for k-mer values below 7 bps to verify compliance with our hypothesis (**Figure 3d**).

These findings are deeply connected to genome size as demonstrated above. Because larger genomes tend to accumulate interspersed repeats, their effective vocabulary expands more slowly (lower β), which algebraically implies a more negative Menzerath exponent. Thus genome length, Heaps compression, and the Menzerath trade-off form a coherent and interrelated triad.

**Evaluation of Zipf's, Zipf-Mandelbrot and truncated power laws for k-mer distributions**

We next examined the previously proposed Zipf's law [10], to model the k-mer distribution profiles across organismal genomes. We find that Zipf's law $f(r) = C/r$, where $C$ is the count of the most frequent k-mer, cannot accurately estimate the decay patterns in k-mer occurrences in the organismal genomes tested (**Figure 4a-c**), consistent with previous reports [11–13]. As exemplified for *Homo sapiens* and *Escherichia coli*, in order for Zipf's law to hold true, the k-mer counts would have to form a straight line with slope of -1 in logarithmic scale; however this pattern does not hold suggesting that a more complex and nuanced model is required to accurately represent the observed distributions (**Figure 4a-b**). The findings are consistent when examining organisms across the taxonomic groups and different k-mer lengths (**Figure 4d-e**). This is quantitatively supported by the $R^2$ values, which remain low across genome assemblies and k-mer lengths, underscoring the inadequacy of Zipf's law in capturing the k-mer frequency - rank relationship in organismal genomes (**Figure 4d**). Finally, consistent patterns emerged across all three superkingdoms and viruses when results were stratified taxonomically (**Figure 4e**). Therefore, we next investigated if other distributions would better fit the k-mer frequency - rank patterns observed in organismal genomes.

We observed that the k-mer frequency distributions exhibit a logarithmic decay, with a heavy-tailed behavior. To model this pattern, we examined known statistical distributions with these characteristics. After testing several candidates on a subset of the data, we selected two that provided the best fit. We focused on the truncated power law and Zipf-Mandelbrot law to model the k-mer frequency - rank relationship across species from the range of taxonomies, because they effectively capture the heavy-tailed nature of genomic k-mer distributions. In both models, the scaling parameter C modulates the overall magnitude of the distribution. The exponent α governs the steepness of the decay: higher α values lead to a steeper drop-off, while lower values yield α more gradual decline (**Figure 5a**). The λ parameter in the truncated power law controls how quickly the tail gets truncated. The β parameter in the Zipf-Mandelbrot controls how sharp the start is and the general shape of the distribution (**Figure 5a**). Higher values of the parameter β horizontally shift the rank axis, so that each item behaves as if it were farther down the list. This flattens the head of the curve and makes the frequency of the top-ranked items decline more gradually. As β values get closer to 0 we approach the standard Zipf's law.

We report particularly strong performance using the proposed models, consistently achieving high goodness-of-fit scores across the range of k-mer lengths analyzed, as estimated with $R^2$ values (**Figure 5a**). For the truncated power law most organisms for k=3 up to k=6 have very small λ values which indicate small to moderate truncation. This means that we observe a power law behavior for small-medium ranks and exponential decay becomes noticeable for larger ranks. Additionally, most organisms also exhibit small α values which indicates an extremely heavy-tailed distribution where most of the mass is concentrated in very high k values. For the Zipf Mandelbrot, most organisms exhibit α values between 2 and 10 with viral data reaching up to 20. This means that there is a steep,rapid power law decay. The β values for all organisms across all k values remain large, resulting in heavy smoothing. The distribution becomes much more uniform for smaller and medium ranks essentially shifting the power law behavior to higher ranks.

We also examined the performance of these models separately in organisms belonging to the different superkingdoms and across viruses. We observe that bacteria and archaea show the best performance across k-mer lengths (**Figure 5b**), however across all k-mer lengths and all taxonomic groups we observe $R^2$ values above 0.91, indicating the fitness of our proposed models is broadly applicable. We provide example cases for *Homo sapiens* and *Escherichia coli*, both of which exhibit high model fit for both of our proposed distributions (**Figure 5c-d**). These observations highlight both the strengths of the Zipf-Mandelbrot and the truncated power laws in modeling k-mer frequencies in organismal genomes.

Lastly, we evaluate the performance of our two distributions specifically for 7-mers across our dataset. We discover that only 79,135 genomes (≈35 %) yield a positive coefficient of determination when fitted with the truncated power law, whereas 146,546 genomes (≈ 65%) return negative $R^2$ values, indicating a systematic failure of this model for the majority of the dataset. In contrast, the Zipf Mandelbrot model achieves a positive fit for virtually every genome examined (only 12 negative fits), underscoring its robustness at this k-mer length **(Supplementary Figure 5)**. Importantly, for the subset of genomes that do fit well with the truncated power law, the estimated parameter λ converges to 0. This vanishing λ implies that, whenever the truncated power law does succeed in fitting the data, it effectively collapses back to an ordinary power-law.

Collectively, these results indicate that as k increases beyond 6 base-pairs, k-mer counts become sparser and more diffusely distributed across ranks, causing the Truncated Power Law to collapse and not be appropriate for estimating the distribution.

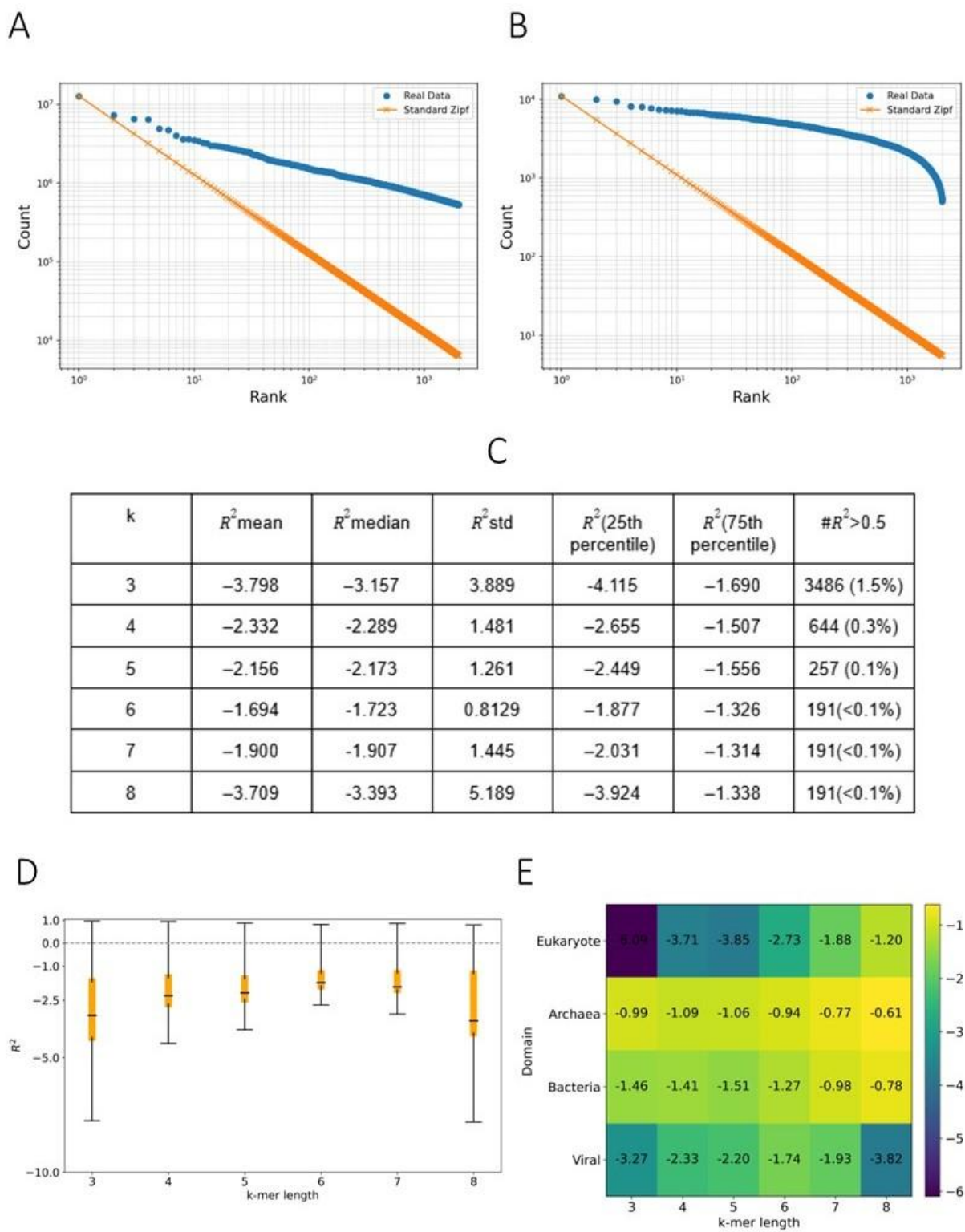

**Figure 4. Evaluation of Zipf's law and power-law models for k-mer frequency distributions.**
**(A)** Observed versus theoretical Zipfian distribution for *Homo sapiens* (k=7), showing a deviation from the expected straight line with slope –1 on a log-log plot, indicating poor fit. **(B)** Similar analysis for *Escherichia coli* (k=6) also demonstrates a lack of adherence to Zipf's law. **(C)** Table showing statistical analysis of the fit of Zipf's law across the entire dataset. **(D)** Box plot showcasing where most of R2 values lie (yellow box indicates 25-75 percentile)**. (E)** Heatmap demonstrating the median value for R2 across taxonomy and k-mer length.

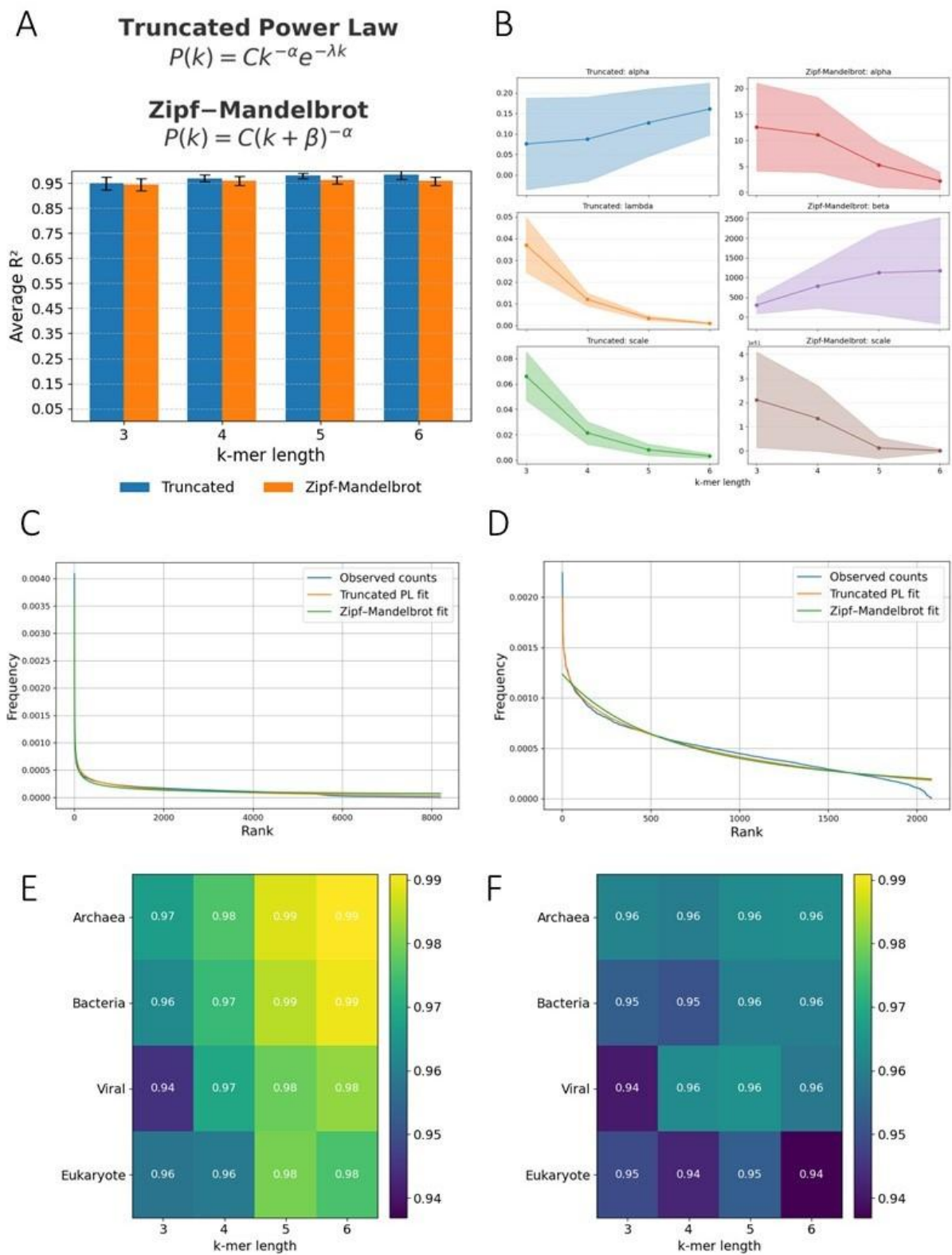

**Figure 5. Zipf-Mandelbrot and truncated power laws model k-mer distributions. (A)** Relationship between the coefficient of determination (R²) for the Zipf-Mandelbrot and the truncated power law models. **(B)** Average fitted parameters for both distributions for all k values.

The error margins represent 1 std above and below the mean. **(C-D)** Model fit for k-mer frequency versus rank in (**C**) *Homo sapiens*, and *Escherichia coli*. **(E-F)** Heatmaps showcasing goodness of fit across superkingdoms and viruses for the Truncated power law (left) and the Zipf Mandelbrot (right) distribution respectively. On all these figures as well as the ones following, 356 organisms were excluded from the truncated power law at k=6 because they showcased heavily negative R2 values that hindered the analysis for the rest. For k=7 where this irregularity strengthens, we perform a separate analysis on **Supplementary Figure 5**.

### Genomic determinants of model fit: the roles of genome size and GC content

To investigate factors influencing the performance of the truncated power law and Zipf-Mandelbrot models, we examined their fit in relation to genomic characteristics, including genome size, genic percentage and GC content. We observed a strong positive correlation between genome size and model performance, with larger genomes exhibiting higher $R^2$ values (**Figure 6a-b; Supplementary figure 6;** Spearman's rank correlation coefficient; Truncated power law, $\rho$ = 0.3846 p-value = 0; Zipf Mandelbrot, $\rho$ = 0.0489 p-value = 0). Smaller genomes often showed degraded fits, as many possible k-mers were absent due to size constraints. Additionally, we found that genomes with moderate GC content (30–60%) had notably lower $R^2$ values compared to those with extreme GC content (**Figure 6c-d; Supplementary figure 7**). This may reflect a more balanced GC content, which leads to less variable k-mer frequencies that are not well captured by heavy-tailed statistical models. We also investigated how the proportion of the genome corresponding to genic regions influences the goodness of fit of the models. For most values of k, the goodness of fit for the truncated power-law model increases as the genic percentage rises (**Figure 6e**). A similar trend is observed for the Zipf-Mandelbrot model (**Figure 6f**). The only exception occurs at k = 3, where the goodness of fit decreases with increasing genic percentage (**Supplementary Figure 8**). For all other k values, higher genic percentages are associated with improved model fit. Together, these results indicate that genome size, GC content, and genic composition significantly influence the suitability of heavy-tailed models, with larger, gene-rich genomes and those with biased GC content providing the best fits.

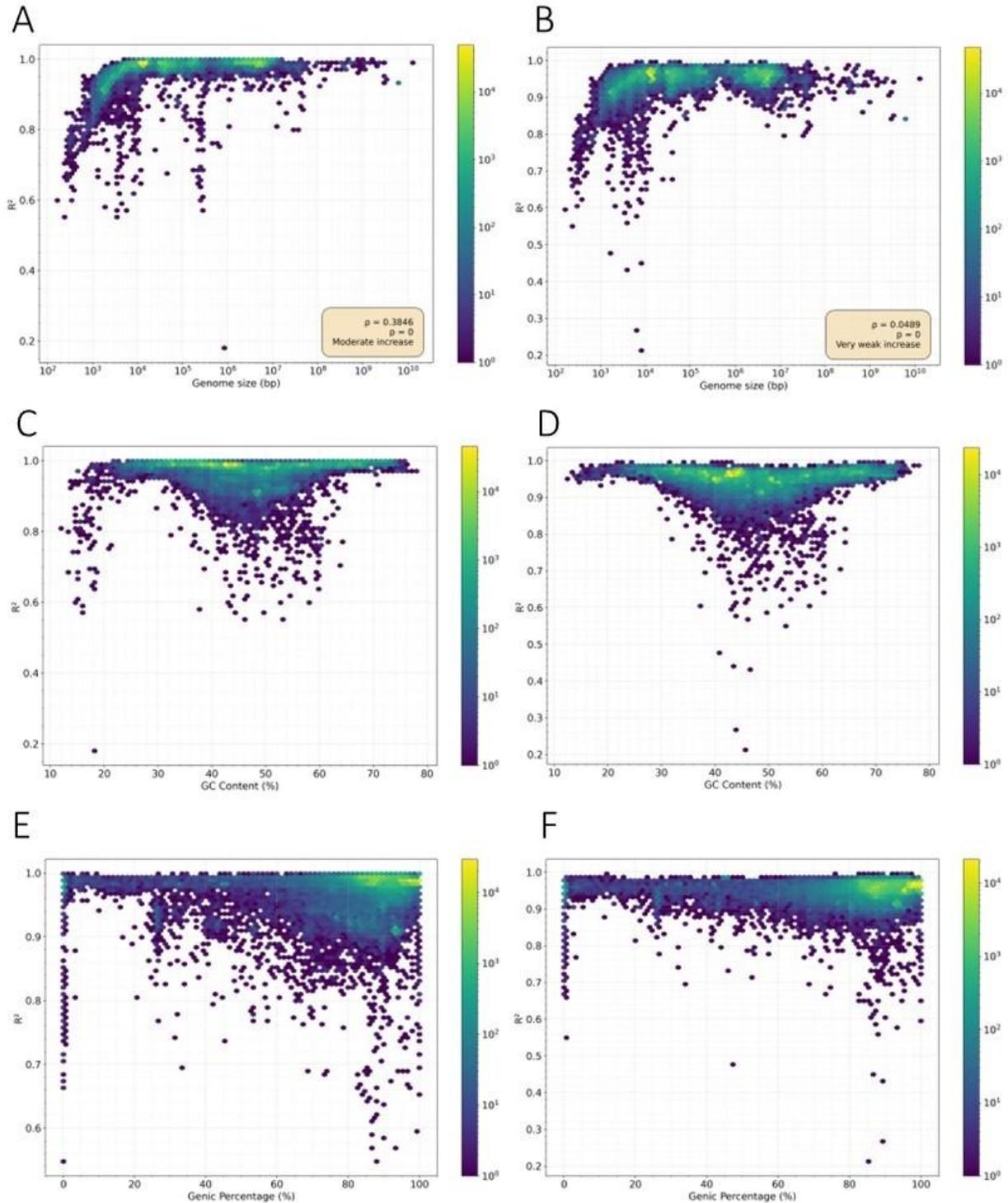

**Figure 6. Comparative performance of Truncated power law and Zipf-Mandelbrot models with different determinants. (A, C, E)** Truncated power law fits for the entire dataset at k=6 in comparison to **(A)** Genome size (log scale), **(C)** GC Content and **(E)** Genic Percentage.**(B, D, F)** Corresponding Zipf-Mandelbrot model fits for the same genomes and determinants.

**Parameter dynamics of heavy-tailed models in organismal and artificial genomes**

To evaluate how synthetic genomic sequences can be compared to their real counterparts, we examined the behavior of the Zipf-Mandelbrot and truncated power-law models across real, shuffled, and synthetic genomes. For the truncated power-law model, organismal genomes display patterns characterized by low α and λ values, indicative of heavy-tailed distributions with minimal exponential cutoff (**Figure 7a**). The Zipf-Mandelbrot model applied to organismal genomes also shows consistently low α values and moderate β values, aside from a few outliers (**Figure 7c**). Shuffled genomes yield α values similar to those of organismal genomes in both the truncated power-law and Zipf-Mandelbrot models. However, for the truncated power law, λ values remain comparable except at k = 7, where they increase sharply, indicating strong exponential truncation. In contrast, the Zipf-Mandelbrot model shows substantially higher β values for shuffled genomes, suggesting that artificial smoothing is required and that the natural heavy-tailed structure is largely disrupted. Next, we constructed synthetic genomes using Evo [22]. Synthetic genomes exhibit a substantial rise in α values and a marked reduction in λ values, approaching zero (**Figure 7b-d**). This shift effectively reduces the truncated model to a pure power law, with α serving as the sole shaping parameter. Synthetic genomes exhibit very low β values and effectively reduce the Zipf-Mandelbrot model to a pure power law, with α values generally higher than those observed in real genomes.

Organismal genomes maintain a heavy-tailed structure. In contrast, synthetic genomes tend to exhibit overly rigid hierarchical organization, often approximating a strict power-law distribution. Shuffled genomes, on the other hand, fail in the opposite direction by disrupting the underlying hierarchy entirely. To approximate any power-law behavior, they require excessive artificial smoothing, as their randomized structure is fundamentally incompatible with the scale-free organization captured by heavy-tailed models. As a result, both synthetic and shuffled genomes deviate markedly from the natural statistical properties observed in real genomes, and both models underperform in these artificial contexts.

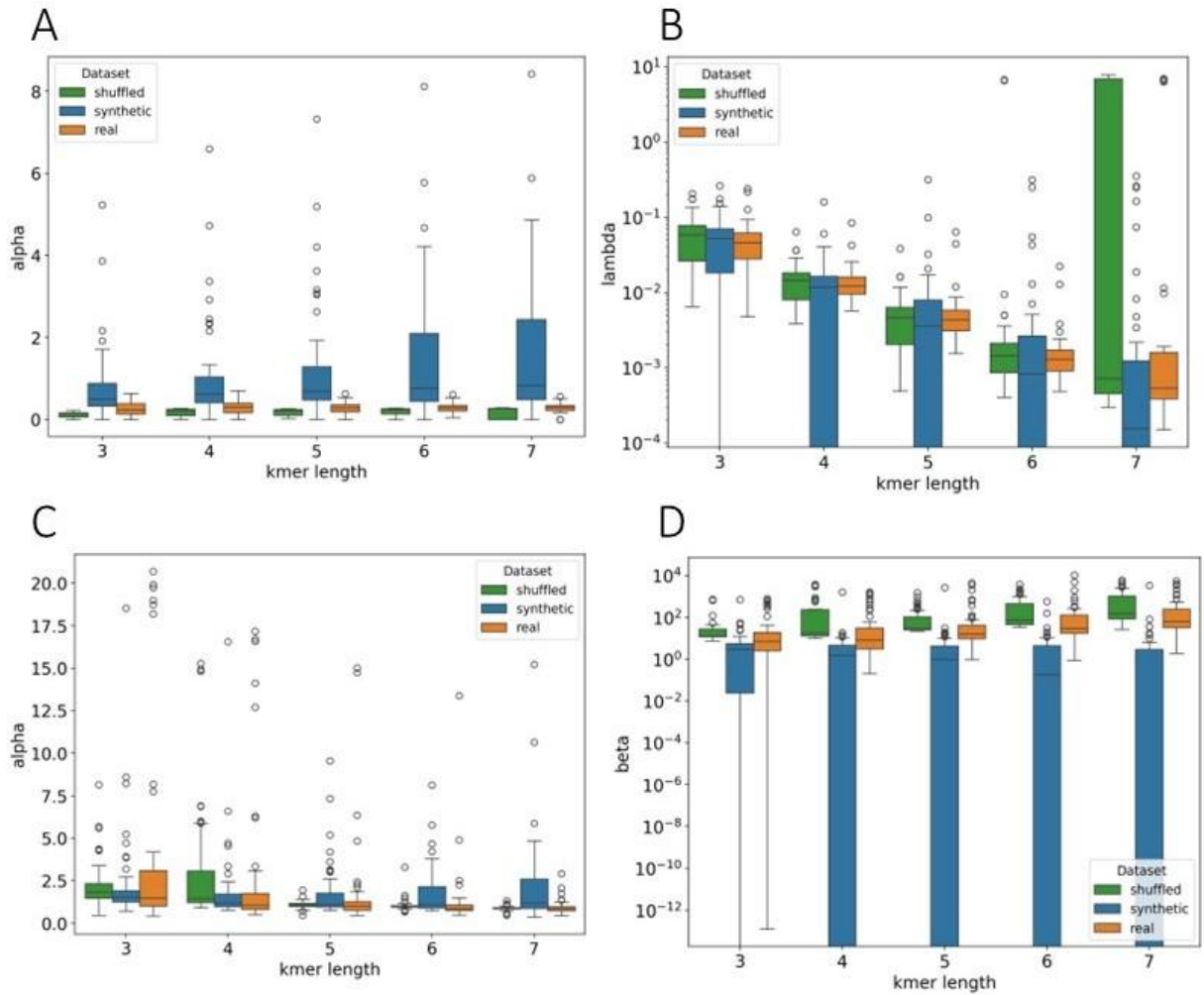

**Figure 7: Box plots of fitted parameters across datasets (synthetic, shuffled and real) and k-mer lengths for truncated power law and Zipf-Mandelbrot models. (A)** illustrates the scaling exponent α and **(B)** λ (log scale) the exponential cutoff parameter for the truncated power law. **(C)** shows the scaling exponent α and **(D)** the shift parameter β (log scale).

**Discussion**

Our large-scale analysis of over 225,000 organismal genome assemblies demonstrates that Heap's law, originally formulated to describe sublinear vocabulary growth in natural language, is applicable to k-mer profiles of organismal genome assemblies. We also find that the classical Zipf's law is statistically inadequate for modeling genomic k-mer frequency distributions, in accordance with other works [8,12]. Despite Zipf's law's success in describing word frequencies in natural language, genomic k-mer profiles exhibit deviations that Zipf's law form cannot capture. In our evaluation of over 225,000 genomes, the observed frequency–rank curves do not follow the ideal Zipfian linear trend (slope ≈ -1 on a log–log plot) expected under Zipf's law. We find that the truncated power law and Zipf-Mandelbrot distributions provide significantly better fits than Zipf's law, for modeling genomic k-mer profiles as both models introduce additional parameters to accommodate deviations from a pure Zipfian decay. We also found that model performance is influenced by genome properties, including genome size, which is positively correlated with fit quality, and GC content which affects the heavy-tailed behavior.

Each distribution approaches our observed data in its own way. The truncated power law treats the entire distribution as fundamentally heavy-tailed and sees small to medium rank behavior as just part of a very shallow power law. The small $\lambda$ values suggest the tail extends quite far before being cut off. The Zipf-Mandelbrot views the data as a steep power law that has been smoothed at small ranks. The large $\beta$ accounts for the non-Zipfian behavior in top ranks (which is what we observed from the start) and the large $\alpha$ values suggest that the process has steep decay. Overall since both distributions provide great fits, we conclude that both models capture the k-mer frequency rankings.

Beyond characterizing natural genomes, our findings have direct implications for synthetic genome benchmarking. Applying our models to synthetic sequences revealed substantial differences between real and simulated genomes. By quantitatively comparing $R^2$ or other fit metrics, we can diagnose deviations in synthetic genomes' k-mer profiles and identify where generative models are not capturing genuine genomic complexity. This approach offers a principled way to benchmark the realism of artificial genomes.

These insights will aid future efforts in modeling genome complexity, in evaluating synthetic sequences, and in understanding organismal genome composition. Future work may also include evaluating this methodology in the protein space and in genomic sub-compartments across taxonomic groups.

## Materials and methods

### Canonical kmers

In this work, in order to account for both the forward and reverse-complement orientations and avoid redundancy, when representing genomic sequences, we use the canonical k-mer form. The canonical representation entails that whenever we parse a k-mer, the reverse complement is also computed and the lexicographically smaller out of the two is kept. For a k-mer of length $k$ (where $k$ is even) to be identical to its own reverse complement, the first $k/2$ bases determine the last $k/2$ bases via complementation. Since we can choose each of the first $k/2$ positions freely from the 4 DNA bases, we get $4^{k/2}$ palindromic k-mers. Thus, the total possible canonical k-mers would be :

$$\text{Canonical } k\text{-mers} = \begin{cases} \frac{4^k + 4^{k/2}}{2} & \text{if } k \text{ is even} \\ \frac{4^k + 2 \cdot 4^{(k-1)/2}}{2} & \text{if } k \text{ is odd} \end{cases}$$

### Modeling k-mer vocabulary growth with Heaps' law

The core of the process involved a sliding window approach to extract k-mers of a specified length, $k$. To account for the double-stranded nature of DNA the canonical form of the k-mers was used. To capture the growth of the vocabulary of unique k-mers as a function of the total number of k-mers, we recorded the cumulative count of distinct k-mers and the total number of k-mers processed at every one hundred evenly spaced intervals that represent 1% of the genome size. After this process was completed, we fitted the Heaps' law $V(n) = K * n^\beta$ where $V$ is the number of unique k-mers, $n$ is the total number of k-mers, and $K$ and $\beta$ are parameters that characterize the vocabulary richness. The fitting process was done using scipy.optimize.curve_fit function which utilizes a non-linear least squares fitting procedure.

### Frequency-rank probability distributions

To examine the count-rank distributions of k-mers across multiple genomes, we first extracted all canonical k-mers from each genome in our dataset. For each genome, we calculated the total occurrences of each k-mer and then sorted these counts in descending order. Next, we converted the raw counts to a normalized frequency by dividing each count with the sum of all k-mer counts for that genome, thus mapping counts into regularized frequencies in the [0,1] interval. This was performed in order to compare k-mer count distributions from thousands of species with widely varying genome sizes and k-mer counts ensuring all distributions are placed on a common scale for meaningful comparisons. By performing multiple tests and reviewing the literature we noticed logarithmic behavior between the counts of the k-mers and their ranks, and more specifically a heavy tailed one. Heavy-tailed distributions are characterized by a relatively slow decay in the tail, leading to a significant portion of the probability mass being contributed by infrequent observations. We used the non-linear least squares routine curve_fit from the scipy library, initializing parameters based on typical positive ranges reported for heavy-tailed distributions. Convergence was assessed by monitoring the fitting procedure's residuals and ensuring the final parameter estimates stabilized across multiple initial guesses. This allowed us to determine which distribution best captured the observed heavy-tailed behavior of k-mer frequencies and to quantify their parameters for subsequent comparative analysis.

**Zipf-Mandelbrot**

Deriving from relevant work regarding the Zipf's law applications across multiple real world domains and its failure to capture genomic kmer count-rank relationships, we tested an extended variation of the Zipf's law, the Zipf Mandelbrot function. Its probability function can be described as:

$$f(k; N, q, s) \ = \ \frac{1}{H_{N,q,s}} \ \cdot \ \frac{1}{(k+q)^s}$$

where k is the rank of the data, $H$ is a regularization term, $s$ is the exponent controlling the decay rate of the distribution and $q$ is the shift parameter that adjusts the distribution for small ranks. Since our data is already normalized and to endure flexibility and freedom we replace the regularization term with a tunable parameter $C$ responsible for scaling the data in the [0,1] interval, thus increasing complexity and precision.

**Truncated power law**

The truncated power law extends the standard power law, by introducing an upper bound beyond which the frequency decays more quickly. It follows :

$$f(k; \alpha, \lambda) = \frac{k^{-\alpha} e^{-\lambda k}}{\int_{k_{\min}}^{\infty} u^{-\alpha} e^{-\lambda u} \, du}$$

where again k is the rank of the data, the integral is an incomplete gamma function that acts as a normalization term, $\alpha$ controls the algebraic decay and $\lambda$ the exponential cutoff rate. We replace the normalization function with another free parameter $C$ controlling the scale. We adopted the truncated power law when modeling heavy-tailed data on a logarithmic scale that exhibit a straight-line segment over an intermediate range but then drop off more rapidly than any pure power law, as is observed in genomic data.

**Metrics**

In order to evaluate the goodness of fit for the presented curves, we use the coefficient of determination, $R^2$. This metric quantifies the proportion of variance in the observed data that is explained by a statistical model. It can be denoted as:

$$R^2 = 1 - \frac{SS_{\text{res}}}{SS_{\text{tot}}}$$

where

$$SS_{\text{res}} = \sum_{i=1}^{n}(y_i - \hat{y}_i)^2 \quad SS_{\text{tot}} = \sum_{i=1}^{n}(y_i - \bar{y})^2$$

and

An $R^2$ value of 1 indicates a perfect fit, while $R^2 = 0$ signifies that the model explains no more variability than a horizontal line. Values below zero occur when predictions are worse than the mean baseline.

**Shuffled genomes**

The shuffling method includes each nucleotide sequence from a compressed FASTA file being randomly shuffled in order to generate a control dataset that preserves nucleotide composition while eliminating previous positional information. Specifically, our method reads each sequence, concatenates its lines into a single string, and randomly permutes the order of its bases using a uniform shuffle. This approach ensures that the overall base composition and sequence length remain unchanged.

**Synthetic genomes**

Synthetic genomes were generated for 50 prokaryotic organisms using the Evo foundation model (Evo-1) [22]. Due to the computational demands of generating extremely long sequences, this study was limited to prokaryotes, as their genomes are substantially smaller than those of eukaryotes. Both the 8,192 and 131,072 context-length base models (evo-1-8k-base and evo-1-131k-base) were employed for sequence generation. For each of the 50 selected organisms, the initial 1,000 bases were extracted from the original genome's FASTA file to serve as a starting prompt for the model. The Evo model then generated the subsequent sequence. This process was conducted iteratively, where the generated output sequence was continually supplied as the new input prompt to predict the next segment of the genome. This iterative generation was repeated until the total length of the synthetic genome matched the length of the original reference genome.

## Code and Data Availability

All relevant code and results can be found on Github at this link: https://github.com/Georgakopoulos-Soares-lab/Investigating-DNA-Words-and-their-Distributions

## Acknowledgements

Research reported in this publication was supported by the National Institute of General Medical Sciences of the National Institutes of Health under award number R35GM155468 and start-up funds awarded to I.G.S.

## Supplementary Material

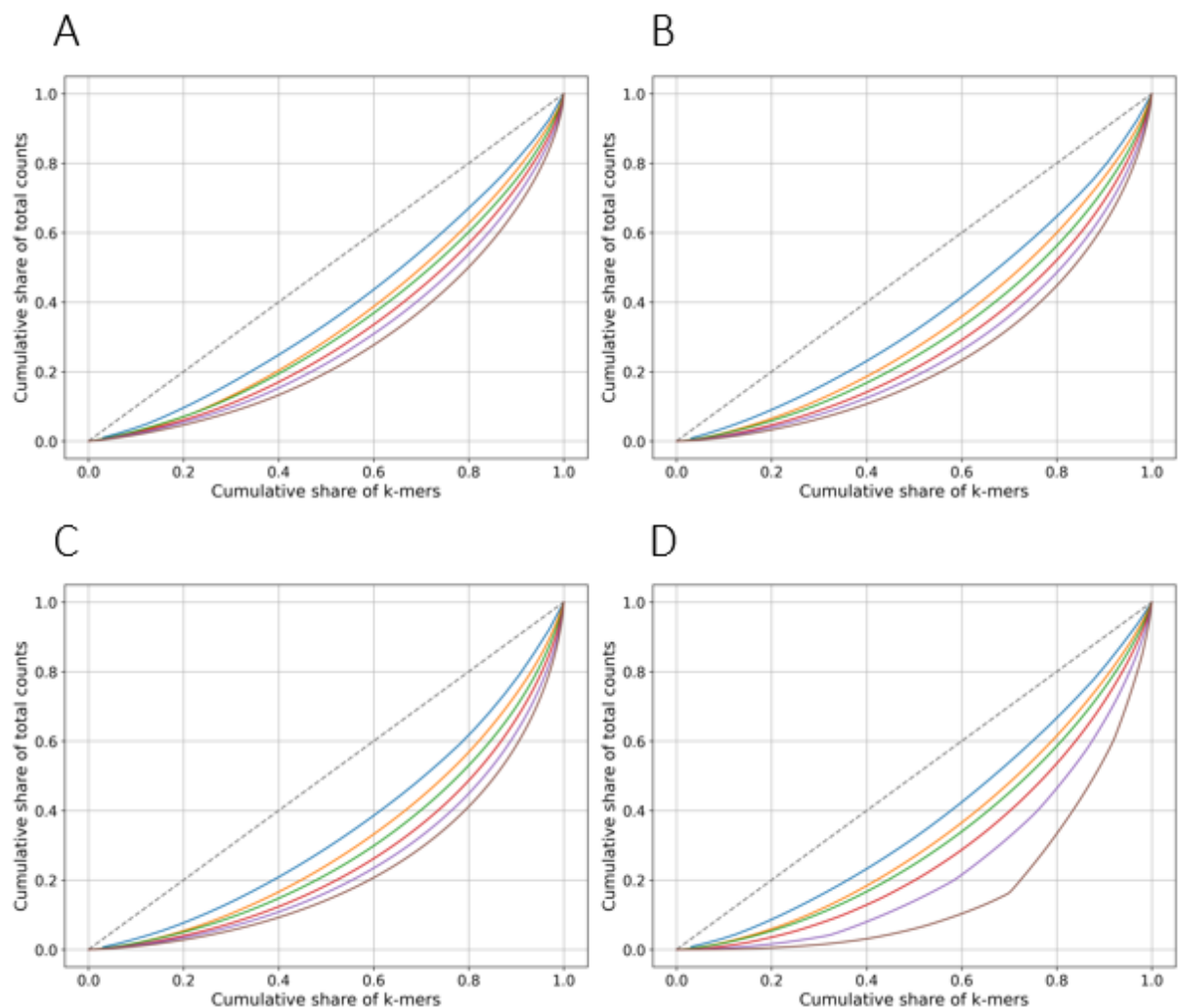

**Supplementary Figure 1.** Lorenz curves illustrate the inequality of k-mer distributions for **A** Eukaryote **B** Bacteria **C** Archaea **D** Viral.

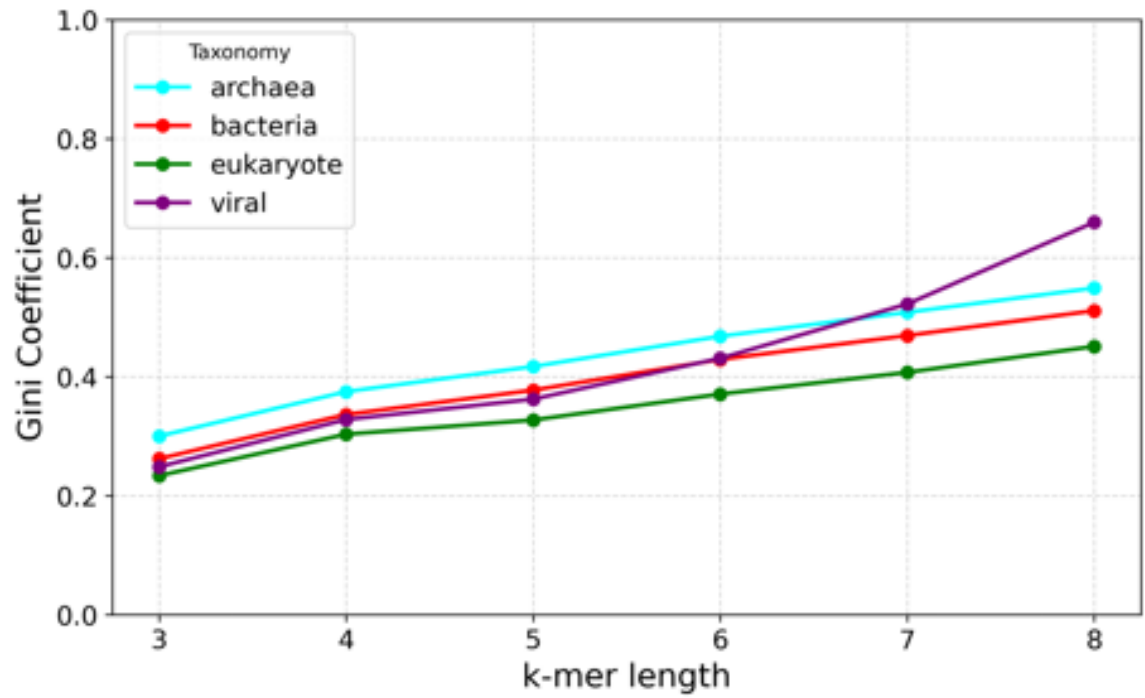

**Supplementary Figure 2:** Gini Coefficient values across taxonomies across k-mer lengths.

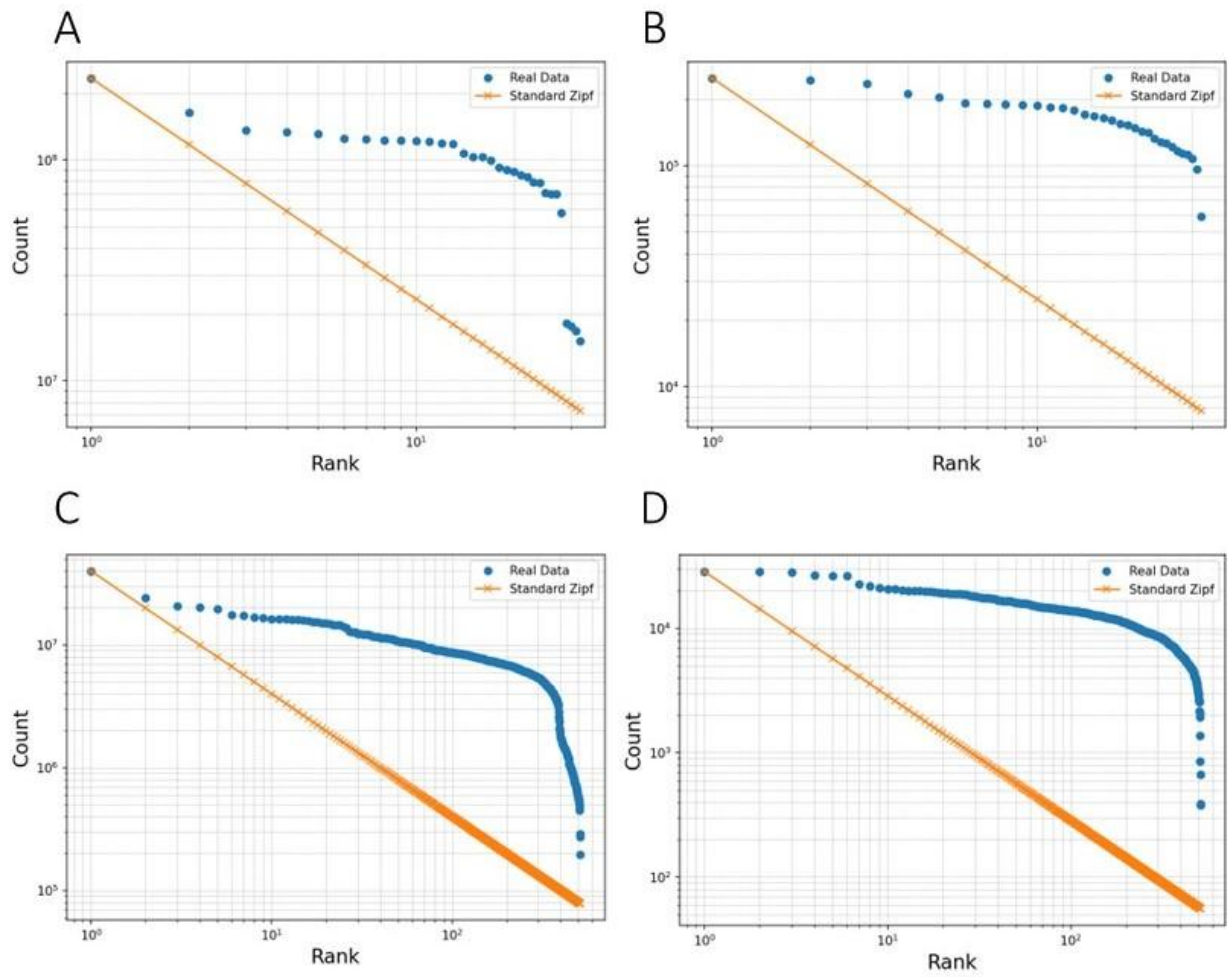

**Supplementary Figure 3.** Observed versus theoretical Zipfian distribution for *Homo sapiens* for k=3 and k=5 **A,C** and for E.coli for k=3 and k=5 **B,D.**

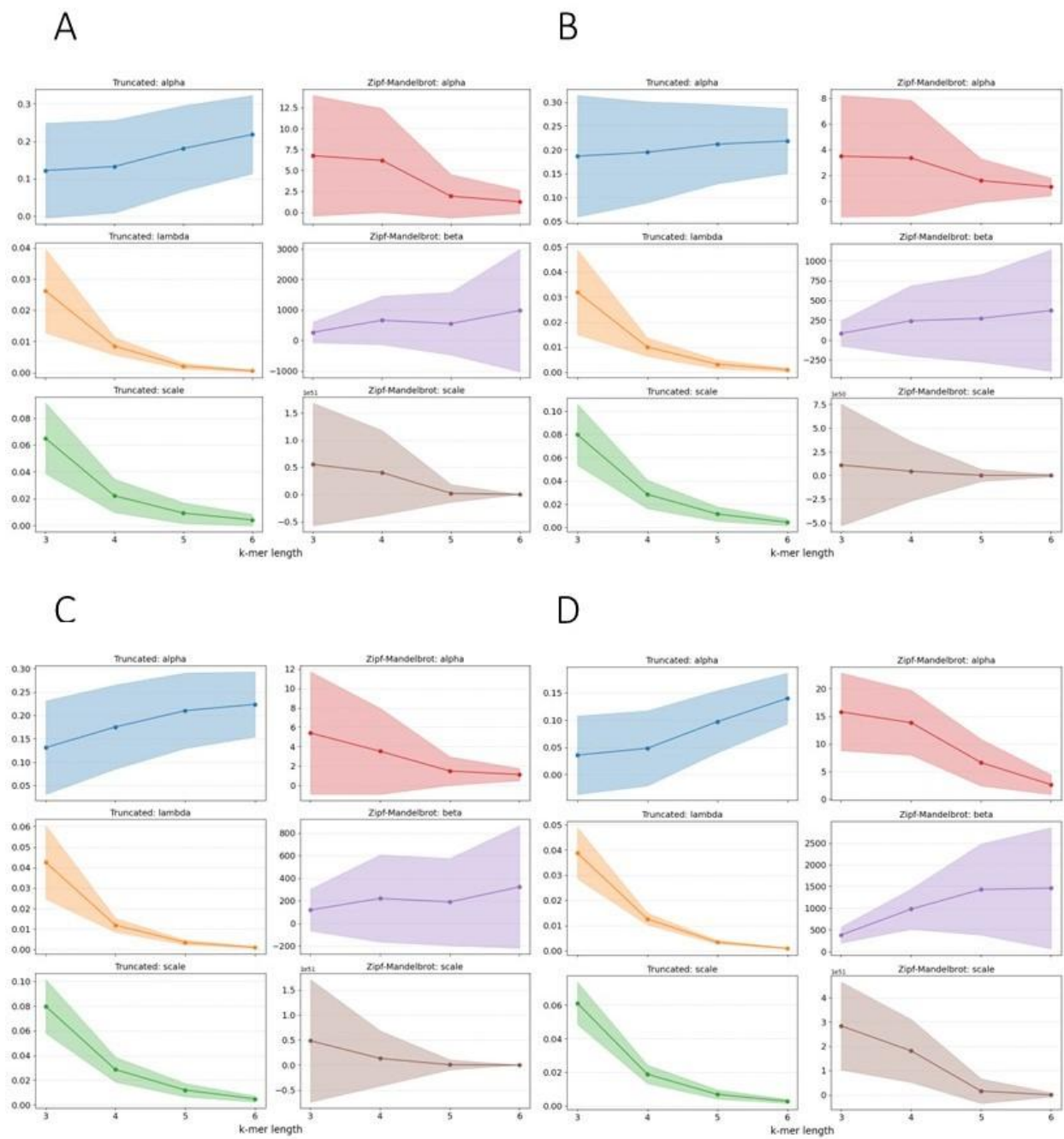

**Supplementary Figure 4:** Average fitted parameters for both distributions for k=3,4,5,6. The error margins represent 1 std above and below the mean in: **A** eukaryote **B** Bacteria **C** archaea **D** viral.

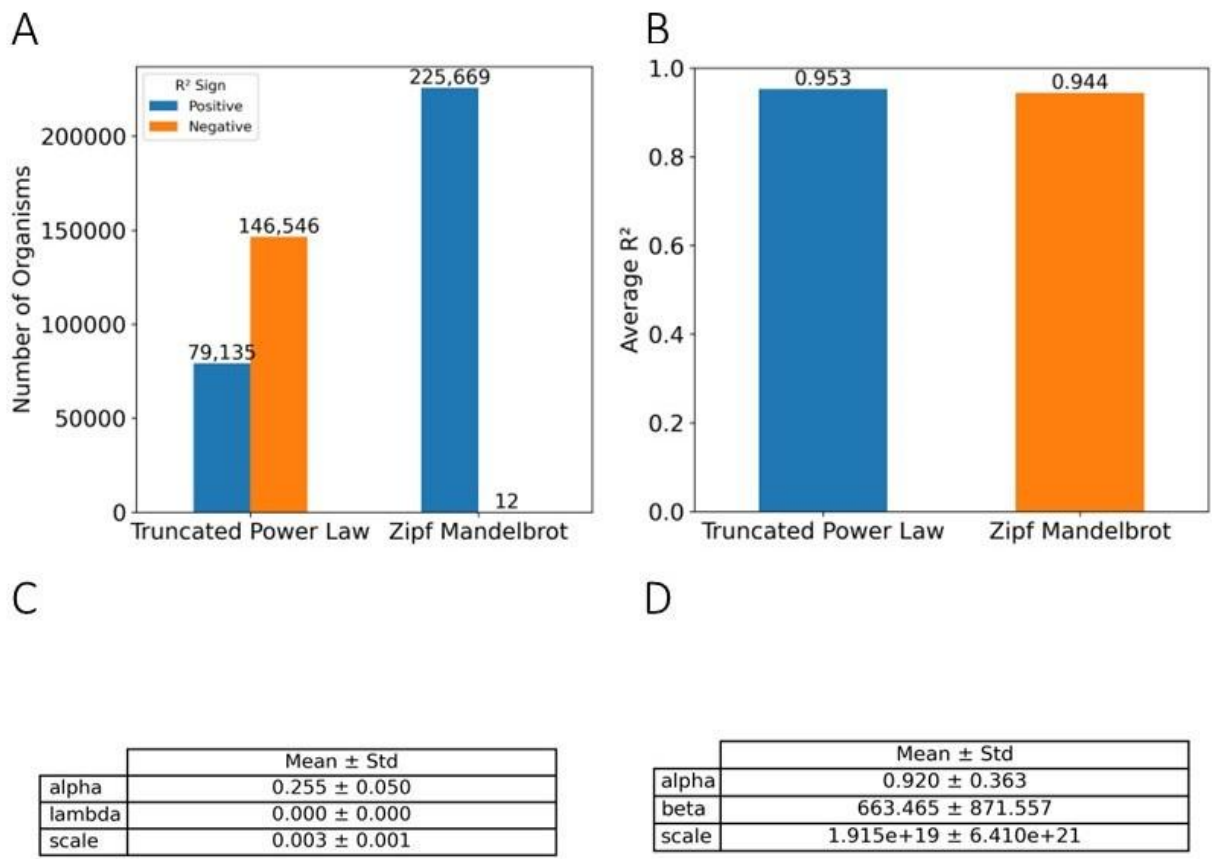

**Supplementary figure 5: Case study made specifically for k-mer length 7. (A)** Depicts the number of organisms showcasing positive and negative R2 values to the truncated power law and Zipf Mandelbrot distributions. **(B)** Shows the mean R2 only for the organisms that exhibited positive R2 values. **(C,D)** Illustrate the mean of all fitted parameters again only for organisms that fitted well with examined curves. **(C)** Truncated power law parameters **(D)** Zipf Mandelbrot parameters.

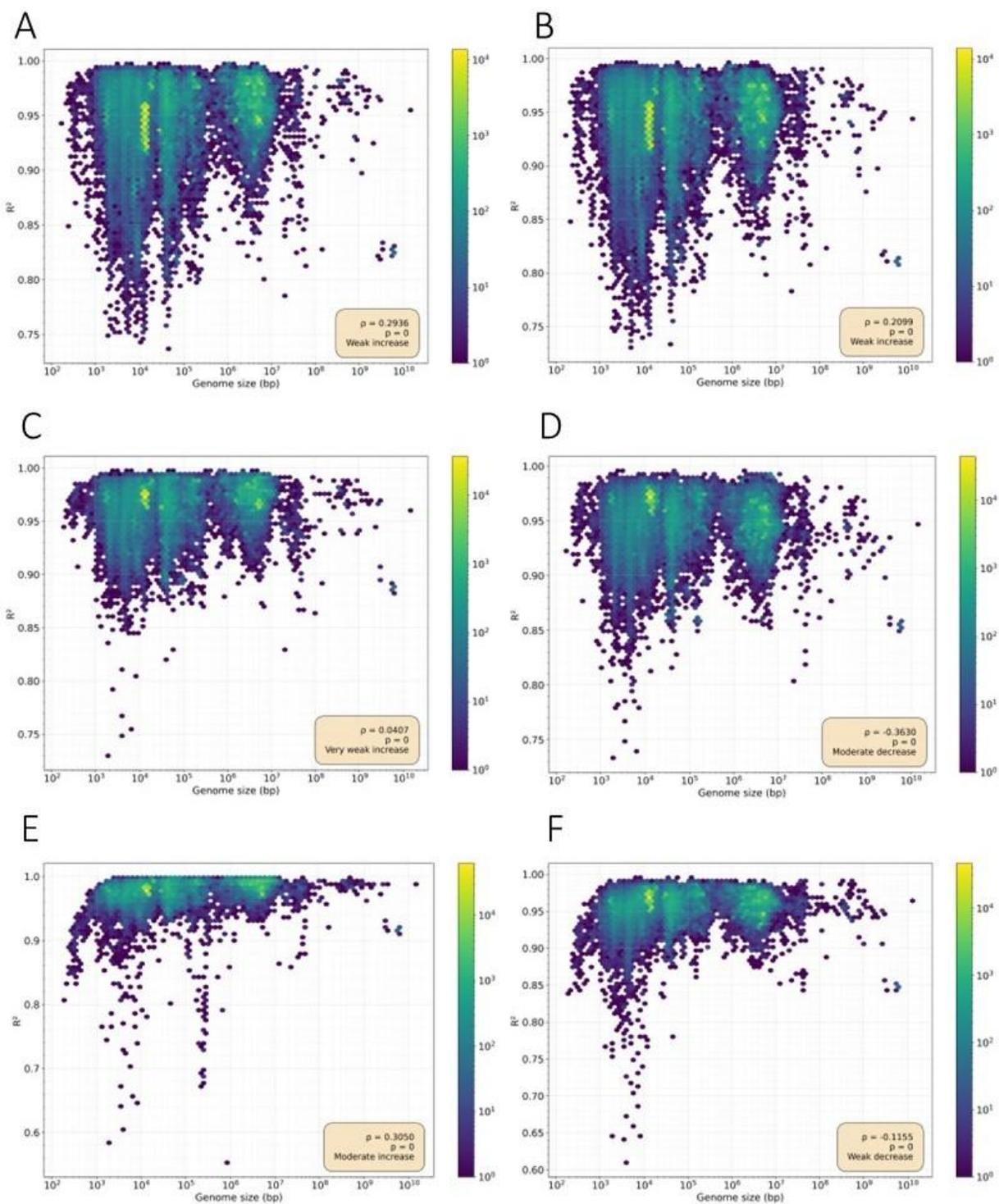

**Supplementary figure 6**: $R^2$ versus Genome size (log scale) for k=3,4,5 for the truncated power law **(A,C,E)** and the Zipf Mandelbrot **(B,D,F)**

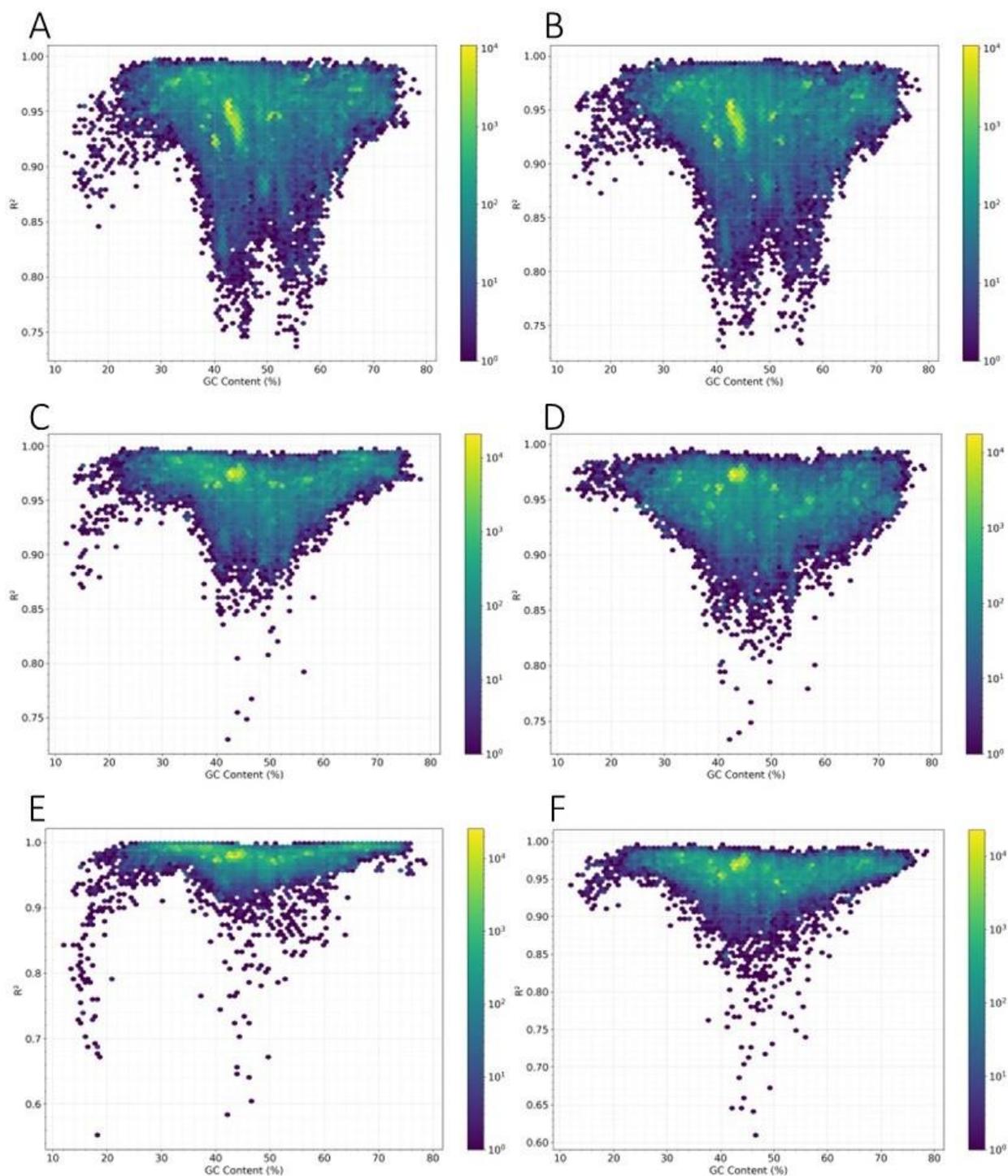

**Supplementary figure 7: $R^2$ versus GC content for k=3,4,5 for both distributions.** The truncated power law results are shown on the left and the Zipf Mandelbrot results on the right.

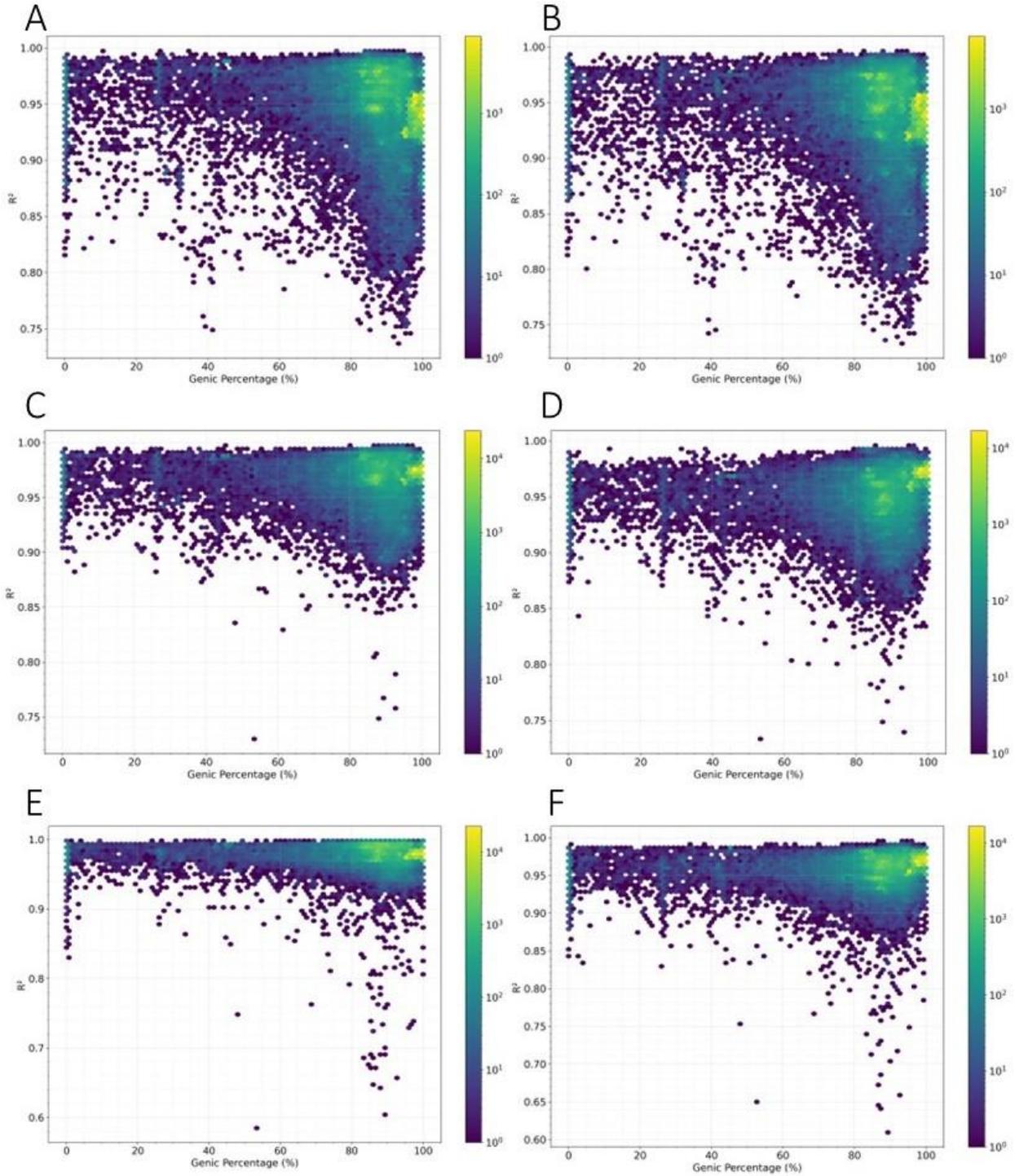

**Supplementary figure 8: $R^2$ versus Genic percentage for k=3,4,5 for both distributions.** The truncated power law results are shown on the left and the Zipf Mandelbrot results on the right.

**References**

1. Chen, F. *et al.* Complete Genome Sequence of Porcine Circovirus 2d Strain GDYX. *Journal of Virology* **86**, 12457 (2012).

2. Fernández, P. *et al.* A 160 Gbp fork fern genome shatters size record for eukaryotes. *iScience* **27**, 109889 (2024).

3. Moeckel, C. *et al.* A survey of k-mer methods and applications in bioinformatics. *Comput Struct Biotechnol J* **23**, 2289–2303 (2024).

4. Yang, Z. *et al.* Intrinsic laws of k-mer spectra of genome sequences and evolution mechanism of genomes. *BMC Evolutionary Biology* **20**, 1–15 (2020).

5. Chor, B., Horn, D., Goldman, N., Levy, Y. & Massingham, T. Genomic DNA k-mer spectra: models and modalities. *Genome Biol* **10**, R108 (2009).

6. Ferrer-I-Cancho, R. & Forns, N. The self-organization of genomes. *Complexity* **15**, 34–36 (2010).

7. Linguistic laws in biology. *Trends in Ecology & Evolution* **37**, 53–66 (2022).

8. Luscombe, N. M., Qian, J., Zhang, Z., Johnson, T. & Gerstein, M. The dominance of the population by a selected few: power-law behaviour applies to a wide variety of genomic properties. *Genome Biol* **3**, RESEARCH0040 (2002).

9. Furusawa, C. & Kaneko, K. Zipf's law in gene expression. *Phys Rev Lett* **90**, 088102 (2003).

10. Mantegna, R. N. *et al.* Linguistic features of noncoding DNA sequences. *Phys. Rev. Lett.* **73**, 3169–3172 (1994).

11. Li, W. Zipf's Law everywhere. *Glottometrics* (2002).

12. Konopka, A. K. & Martindale, C. Noncoding DNA, Zipf's law, and language. *Science* **268**, 789 (1995).

13. Chatzidimitriou-Dreismann, C. A., Streffer, R. M. F. & Larhammar, D. Lack of Biological Significance in the 'Linguistic Features' of Noncoding DNA—A Quantitative Analysis.

*Nucleic Acids Res* **24**, 1676–1681 (1996).

14. Sheinman, M., Ramisch, A., Massip, F. & Arndt, P. F. Evolutionary dynamics of selfish DNA explains the abundance distribution of genomic subsequences. *Scientific Reports* **6**, 1–8 (2016).

15. Greg Warr, L. H. The Architecture of the Genome Integrates Scale Independence with Inverse Symmetry. *Academia Molecular Biology and Genomics* (2025) doi:10.20935/AcadMolBioGen7650.

16. Li, W. Menzerath's law at the gene-exon level in the human genome. *Complexity* **17**, 49–53 (2012).

17. Chacoma, A. & Zanette, D. H. Heaps' Law and Heaps functions in tagged texts: evidences of their linguistic relevance. *R Soc Open Sci* **7**, 200008 (2020).

18. Tettelin, H., Riley, D., Cattuto, C. & Medini, D. Comparative genomics: the bacterial pan-genome. *Curr Opin Microbiol* **11**, 472–477 (2008).

19. Bonnie, J. K., Ahmed, O. Y. & Langmead, B. DandD: Efficient measurement of sequence growth and similarity. *iScience* **27**, 109054 (2024).

20. Range-limited Heaps' law for functional DNA words in the human genome. *Journal of Theoretical Biology* **592**, 111878 (2024).

21. Baeza-Yates, R., Glaz, J., Gzyl, H., Hüsler, J. & Palacios, J. L. *Recent Advances in Applied Probability*. (Springer Science & Business Media, 2006).

22. Nguyen, E. *et al.* Sequence modeling and design from molecular to genome scale with Evo. *Science* (2024) doi:10.1126/science.ado9336.